\documentclass[aps,prl,twocolumn]{revtex4}
\usepackage{amssymb}

\def\sfF{{\sf F}}
\def\sfT{{\sf T}}

\begin{document}
\title{Fundamental Limits of ``Ankylography'' due to Dimensional Deficiency}
\author{Haiqing Wei}
\affiliation{oLambda, Inc., Los Altos, CA 94022\\
E-mail: davidhwei@yahoo.com}
\begin{abstract}
Single-shot diffractive imaging of truly 3D structures suffers from a dimensional deficiency and does not scale. The applicability of ``ankylography'' is limited to objects that are small-sized in at least one dimension or that are essentially 2D otherwise.
\end{abstract}
\maketitle

Raines {\em et al}. \cite{Raines10} proposes a method, dubbed ``ankylography'', for three-dimensional (3D) structure determination using single-shot diffractive imaging (SSDI). But the conclusion without limitation that the 3D structure of an object is ``in principle encoded into a 2D (two-dimensional) diffraction pattern on the Ewald sphere'' and may be inverted by SSDI is inadequately substantiated and conceptually misleading. Here we point out that SSDI in general suffers from a dimensional deficiency in using a 2D observation with a quadratically growing number of degrees of freedom (NDF) to recover a 3D structure with a cubically growing number of unknowns, when the size of a genuine 3D object increases. Practically obtainable signal-to-noise ratio (SNR) and measurement accuracy limit the applicability of ``ankylography'' to objects that are small-sized in at least one dimension or that are essentially 2D otherwise.

The rate of reliable information transfer via a spatial or temporal channel is fundamentally limited by the channel capacity that is determined by the NDF available therein and obtainable SNR \cite{Gallager68}. The resolving powers of telescopes or radio antennas and microscopes including SSDI are all limited in much the same manner. A steep (exponential) price in signal power has to be paid to obtain data rates or resolutions significantly beyond that are supported by the available NDF \cite{Gallager68,Hansen81}. SSDI uses a spatial channel characterized by a bounded linear operator $\sfT:L^2([0,l]^3)\!\rightarrow\! L^2(S)$ that projects a real space amplitude with support $[0,l]^3$ onto the Ewald sphere $S=\{(f,g,h):f^2+g^2+h^2=1\}$, where $l$ is measured in units of the wavelength, $(f,g,h)$ represents a normalized spatial frequency. Well known is the existence of a pair of orthonormal bases $\{u_i\}_{i\ge 1}$ and $\{v_i\}_{i\ge 1}$, called normal modes, and the associated modal gains $\{\lambda_i(\sfT)\}_{i\ge 1}$, being all positive and arranged in a nondecreasing order, such that $\sfT u_i=\sqrt{\lambda_i}v_i$, $\forall\,i\ge 1$ \cite{Gallager68,Miller98}. For any modal cutoff threshold $\epsilon\in(0,1)$ as determined by practically obtainable SNR and accuracy in signal measurement, $N(\sfT,\epsilon)=\max\{\,i:\lambda_i(\sfT)\ge\epsilon\}$ is the number of usable normal modes. We shall prove that when $l$ is large,
\begin{equation}
N(\sfT,\epsilon)\le 8l^2+O(|\log\epsilon|\,l\log l+\log^2\epsilon\log^2l),
\label{Nbound1}
\end{equation}
which grows far too slow in comparison with the number of unknowns $O(l^3)$ in the structure of a general 3D object. The dearth of NDF would persist even if $1/\epsilon$ grew exponentially as $l$ increased, so long as the exponent grew no faster than $l^{3/2}$, more specifically, $|\log\epsilon|=O(l^{1.5-\delta})$, for any fixed small $\delta>0$.

By considering the distribution of received energy between two hemispheres, it is easy to see that $N(\sfT,\epsilon)\le N(\sfT_+,\epsilon/2)+N(\sfT_-,\epsilon/2)\le 2N(\sfT_{\scriptscriptstyle\square},\epsilon/2)$, where $\sfT_{\pm}:L^2([0,l]^3)\!\rightarrow\! L^2(S_{\pm})$ and $\sfT_{\scriptscriptstyle\square}:L^2([0,l]^3)\!\rightarrow\!
L^2(S_{\scriptscriptstyle\square})$ are restrictions and extension of $\sfT$ to the corresponding codomains $S_{\pm}=\{(f,g,h):f^2+g^2+h^2=1,\,h\gtrless 0\}$ and $S_{\scriptscriptstyle\square}=\{(f,g):|f|\le 1,\,|g|\le 1\}=[-1,1]^2$. $\sfT_{\scriptscriptstyle\square}=\sfT_z\sfF_y\sfF_x$ is a product of three linear operators, with
$$[\sfF_y\sfF_x\rho](f,g,z)=\frac{1}{4l}\int_0^l\!\!\int_0^l
\rho(x,y,z)e^{-i2\pi(fx+gy)}dxdy$$
being a bounded linear operator of norm $1$, and
$$[\sfT_z\sigma](f,g)=\frac{1}{\sqrt{l}}\int_0^l
\sigma(f,g,z)e^{-i2\pi[(1-f^2-g^2)^{1/2}-1]z}dz$$
being also a bounded linear operator of unit $L^2$ operator norm \cite{Rynne08}. Well known results for operators of ``time and frequency limiting'' \cite{Landau80} state that
\begin{equation}
N(\sfF_x,\epsilon)=N(\sfF_y,\epsilon)=2l+O(|\log\epsilon|\log l),
\end{equation}
and the $x$-$y$ separability implies that
\begin{eqnarray}
N(\sfF_y\sfF_x,\epsilon)&\ge&N(\sfF_y,\epsilon^{1/2})N(\sfF_x,\epsilon^{1/2}),\\
N(\sfF_y\sfF_x,\epsilon)&\le&N(\sfF_y,\epsilon)N(\sfF_x,\epsilon),
\end{eqnarray}
consequently,
\begin{eqnarray}
&&N(\sfF_y\sfF_x,\epsilon)\,=\,[2l+O(|\log\epsilon|\log l)]^2\nonumber\\
&=&4l^2+O(|\log\epsilon|\,l\log l+\log^2\epsilon\log^2l).
\end{eqnarray}
However, the $\sfT_z$ term makes the singular value distribution of $\sfT_{\scriptscriptstyle\square}$ more complicated. Fortunately, an operator inequality comes to the rescue \cite{Bhatia07}. It follows from
\begin{equation}
\sfT_{\scriptscriptstyle\square}^*\sfT_{\scriptscriptstyle\square}\le
\|\sfT_z\|^2_2\,\,\sfF_x^*\sfF_y^*\sfF_y\sfF_x=\sfF_x^*\sfF_y^*\sfF_y\sfF_x
\end{equation}
that $\lambda_i(\sfT_{\scriptscriptstyle\square})\le\lambda_i(\sfF_y\sfF_x)$, $\forall\,i\ge 1$. It is now obvious that
\begin{eqnarray}
&&N(\sfT,\epsilon)\,\le\,2N(\sfT_{\scriptscriptstyle\square},\epsilon/2)\le 2N(\sfF_y\sfF_x,\epsilon/2)\nonumber\\
&=&8l^2+O(|\log\epsilon|\,l\log l+\log^2\epsilon\log^2l),\label{Nbound2}
\end{eqnarray}
which constitutes a rigorous proof of equation (\ref{Nbound1}).

It is worth noting the fundamental nature of the limitation, that single-shot diffraction does not convey sufficient information to invert the 3D structure of an object, even the amplitude (instead of intensity) of the diffracted field is sampled continuously and measured directly with no phase ambiguity. With practically obtainable SNR and measurement accuracy that determine a threshold $\epsilon$ of modal cutoff, any signal in the linear space spanned by the normal modes of orders higher than $N(\sfT,\epsilon)$ is essentially lost in transmission or attenuated beyond detection. Oversampling and inversion algorithms are irrelevant in this context. Indeed, the finiteness of an object ensures that the entire diffraction field is uniquely determined by a finite number of Nyquist-sampled amplitude values. With a limited NDF, the only way to transmit and retrieve more information is to increase the number of significant figures in the measured amplitudes, which however quickly becomes prohibitively expensive.

In summary, SSDI of truly 3D structures does not scale. The applicability of ``ankylography'' is limited to objects that are small-sized with respect to the wavelength in at least one dimension or have structures being essentially 2D in complexity. That may be the case in Raines {\em et al}.'s computer tests and preliminary experiment \cite{Raines10}. Raines {\em et al}. \cite{Raines10} also put much emphasis on certain ``physical constraints'', many of which as being described and used are actually steps of numerical procedures instead of mathematical constraints of model formulation, while the bona fide physical constraints of continuity and boundedness in the support and uniformity outside are automatically satisfied by the normal modes in the present formulation. The nonnegativity of the object field fixes only a single degree of freedom, {\it i.e.}, a global level shift. The incorporation of more physical constraints, to the extreme of having the majority of the $O(l^3)$ unknowns fixed, could arguably alleviate the problem of dimensional deficiency, however that diminishes the generality and appeal of ``ankylography''.

\end{document}